\documentclass{article}
\usepackage{spconf,amsmath,graphicx}
\usepackage{amsfonts}
\usepackage{multirow}
\usepackage{cite}
\usepackage{svg}
\usepackage{url}
\usepackage{xcolor}

\usepackage{hyperref}

\title{A KNOWLEDGE DISTILLATION FRAMEWORK FOR ENHANCING EAR-EEG BASED SLEEP STAGING WITH SCALP-EEG DATA}
\name{\begin{tabular}{c} Mithunjha Anandakumar$^{\star \dagger \ddagger}$, Jathurshan Pradeepkumar$^{\star \dagger \ddagger}$,\\Simon L. Kappel$^{ \diamond}$,   Chamira U. S. Edussooriya$^{\dagger \mathsection}$, Anjula C. De Silva$^{ \dagger}$\thanks{$^{\star}$ These authors contributed equally to the work.}\end{tabular} \vspace{-0.2cm}}

\address{$^{\dagger}${Dept. of Electronic and Telecommunication Engineering, University of Moratuwa, Sri Lanka} \\ $^{\ddagger}$Faculty of Arts and Sciences, Harvard University, USA \\ $^{\diamond}$Dept. of Electrical and Computer Engineering, Aarhus University, Denmark \\ $^{\mathsection}$ Dept. of Electrical and Computer Engineering, Florida International University, USA \vspace{-0.5cm}}

\begin{document}
%
\maketitle
\begin{abstract}

Sleep plays a crucial role in the well-being of human lives. Traditional sleep studies using Polysomnography are associated with discomfort and often lower sleep quality caused by the acquisition setup. Previous works have focused on developing less obtrusive methods to conduct high-quality sleep studies, and ear-EEG is among popular alternatives. However, the performance of sleep staging based on ear-EEG is still inferior to scalp-EEG based sleep staging. In order to address the performance gap between scalp-EEG and ear-EEG based sleep staging, we propose a cross-modal knowledge distillation strategy \footnote[1]{\url{https://github.com/Mithunjha/EarEEG_KnowledgeDistillation}}, which is a domain adaptation approach. Our experiments and analysis validate the effectiveness of the proposed approach with existing architectures, where it enhances the accuracy of the ear-EEG based sleep staging by $3.46\%$ and Cohen’s kappa coefficient by a margin of $0.038$.

%
\end{abstract}


%

\vspace{-2mm}
\section{Introduction}
\label{sec:intro}

Humans spend one-third of their lifetime in sleep, and decades of research have underlined the negative effects of poor sleep on mental and physical well-being. Polysomnography (PSG) is a comprehensive and the gold-standard study currently used in clinics to assess the quality of sleep and to diagnose sleep related disorders. In PSG, several physiological signals such as; electroencephalogram (EEG), electrocardiogram (ECG), electromyogram (EMG), electrooculogram (EOG), and SpO2 are recorded followed by manual annotation of sleep stages by sleep experts based on standard guidelines such as Rechtschaffen and Kales (R\&K) \cite{rechtschaffen1968manual} or American Academy of Sleep Medicine (AASM)\cite{Iber2007}, to aid sleep stage detection and sleep disorder diagnosis.

PSG enables high-quality sleep assessment however, the signal acquisition setup is obstructive to sleep since the patient has to wear multiple sensors and electrodes. The quality of patients’ sleep is compromised due to the discomfort caused by these sensor systems and the unfamiliarity of the hospital environment, thus affecting the diagnostic quality of the study, and leading to treatment errors. Additionally, PSG is a complex and expensive setup that requires expert assistance and specialized laboratories. Therefore, sleep studies are generally conducted at a sleep clinic, which limits PSG from being used in home-based and long-term settings. 

In order to overcome the aforementioned limitations in sleep studies, there have been several efforts to build a simple and comfortable system for high-quality home-based sleep monitoring. Previous works have focused on utilizing either one or combinations of physiological signals such as ear-EEG \cite{nakamura2019hearables, nakamura2017automatic, mikkelsen2017automatic, mikkelsen2019machine,tabar2021ear, mikkelsen2019accurate }, ECG \cite{wei2019multi}, respiratory signal \cite{sun2019hierarchical}, EOG \cite{sun2019hierarchical}, heart rate \cite{zhang2018sleep}, photoplethysmography\cite{korkalainen2020deep}, and actigraphy \cite{kapella2017actigraphy,zhang2018sleep} for sleep monitoring.

Ear-EEG based sleep monitoring has advantages in terms of enhanced comfort and portability when compared to a full PSG montage. This enables long-term sleep monitoring in home settings. Ear-EEG is a potential alternative for PSG, since there is a good correspondence between hypnograms measured with ear-EEG and scalp-EEG \cite{kidmose_2013, mikkelsen2017keyhole}. Several studies have strongly validated that the ear-EEG recordings contain the necessary information to classify sleep stages with an acceptable level of accuracy \cite{mikkelsen2019accurate, mikkelsen2017automatic, nakamura2019hearables, nakamura2017automatic, tabar2021ear, mikkelsen2019machine}. Manually annotating sleep stages on the recordings is a labor-intensive and time-consuming process, hence past research has focused on automating the process \cite{pradeepkumar2022towards,usleep,xsleepnet,phan2022sleeptransformer}. Similar techniques have been explored to classify sleep stages from ear-EEG \cite{mikkelsen2017automatic,mikkelsen2019machine}. However, the performance of sleep staging based on ear-EEG is still inferior to  scalp-EEG based sleep staging even though there is high mutual information between ear-EEG and scalp-EEG. The performance gap could be attributed to the discrepancy in signal-to-noise ratio due to spatial differences in the signal acquisition site and poorer spatial resolution of ear-EEG \cite{mikkelsen2017keyhole}. The unavailability of large publicly available ear-EEG datasets makes this task more challenging. In our study, we focus on improving the sleep staging performance of existing deep-learning based algorithms on ear-EEG by leveraging effective training strategies and knowledge from scalp-EEG recordings. 


In our work, we employ \emph{cross-modal knowledge distillation}, a domain adaptation technique to enhance the performance of ear-EEG based sleep staging. Knowledge distillation \cite{hinton2015distilling} has been explored in the scalp-EEG domain for various applications such as emotion recognition \cite{zhang2022visual}, improving ECG based sleep staging\cite{joshi2021deep}, etc.  Since there is high mutual information between ear-EEG and scalp-EEG\cite{mikkelsen2017keyhole}, we hypothesize that the performance of ear-EEG based sleep staging can be improved by forcing the model to learn a similar feature representation from scalp-EEG data, such that the common features in both domains can be extracted. This should enable the model to better extract important features contained in the ear-EEG data for better sleep staging. To the best of our knowledge, this is the \textit{first} of such works focusing on enhancing ear-EEG based sleep staging by employing knowledge distillation strategies and opens up an unexplored domain of techniques to enhance ear-EEG based sleep staging.

\vspace{-1.2em}
\section{Methodology}
\label{sec:method}

\begin{figure}[t!]
    \vspace{-1.5em}
    \centering
    \includegraphics[width = 0.9\linewidth]{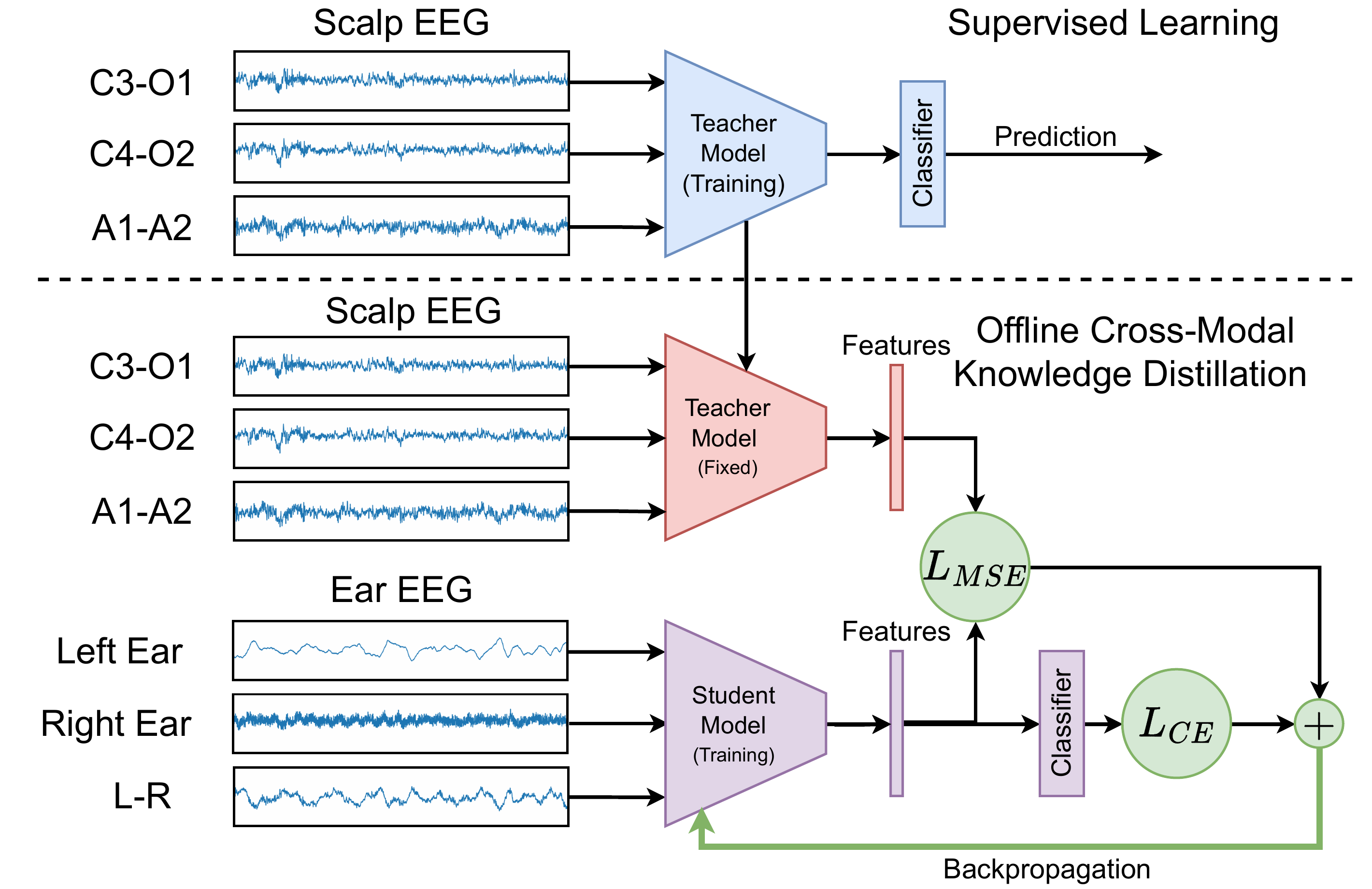}
    \vspace{-1em}    
    \caption{Offline cross-modal knowledge distillation framework}
    \label{fig:offline}
    \vspace{-1.32em}
\end{figure}

\subsection{Problem Definition}
\label{subsec:problem}

In this paper, we address the problem of cross-domain sleep stage classification to enhance ear-EEG based sleep staging. The dataset used in the study consists of simultaneously collected scalp-EEG and ear-EEG recordings \cite{mikkelsen2017automatic}. Our training set, with the size of $N$, consists of labeled and paired $30$ s epochs of scalp-EEG and ear-EEG signals $\{x^s_{i},x^t_{i},y_{i}\}_{n=1}^{N}$, where $(x^s_{i},x^t_{i},y_{i})$ $\in$ $\mathcal{X}^s\times\mathcal{X}^t\times\mathcal{Y}$. Here, $\mathcal{X}^s,\mathcal{X}^t$ $\in$ $\mathbb{R}^{T \times C}$ are the input space of the recorded signals, where $\mathcal{X}^s$ is the source domain of scalp-EEG signals, and $\mathcal{X}^t$ is the target domain of ear-EEG signals. $T$ and $C$ represent the time steps in a $30$ s epoch and the recorded number of channels, respectively. $\mathcal{Y}$ $\in$ $\mathbb{R}^{K}$ denotes the output space of sleep stages, where $K$ represents the number of sleep stages considered. Based on AASM\cite{Iber2007}, we consider $K = 5$, such that $\mathcal{Y}$ $\in$ \{WAKE, N1, N2, N3, REM\}. Under a cross-domain setting, our goal is to learn a function $f_\theta: \mathcal{X}^t \xrightarrow{} \mathcal{Y}$ by utilizing the knowledge in the $\mathcal{X}^s$ and known $\mathcal{Y}$. This is achieved by minimizing the error $E=\mathbb{E}_{(x^t,x^s,y)}(\mathcal{L}(f_\theta(x_i^t),x_i^s,y_i)$ on the given training dataset. Here $\mathcal{L}$ denotes a loss function.


\vspace{-0.9em}
\subsection{Knowledge Distillation Framework}
Knowledge distillation
\label{subsec:KD}
\cite{hinton2015distilling} is a domain adaptation technique used to transfer knowledge from a teacher model to a student model. In knowledge distillation, the student model learns to mimic the teacher model by leveraging the embedded knowledge to achieve similar or higher performance. In our study, we employ cross-modal knowledge distillation. During training, the teacher model is only exposed to the source domain of scalp-EEG signals. Then, the embedded knowledge is distilled to the student model, which is trained in the target domain of ear-EEG signals.

There are different types of knowledge distillation approaches \cite{gou2021knowledge} such as response-based, feature-based, and relation-based knowledge distillation. We focused on improving ear-EEG based sleep staging by forcing the model to learn a similar feature representation to scalp-EEG, such that the common features in both domains could be extracted. This enabled the model to extract important features buried in the ear-EEG signals for better sleep staging. To test and validate our approach, feature-based knowledge distillation\cite{gou2021knowledge} was employed under two different strategies; 1) offline, and 2) online knowledge distillation.

\begin{figure}[t]
    \centering
    \includegraphics[width = 0.9 \linewidth]{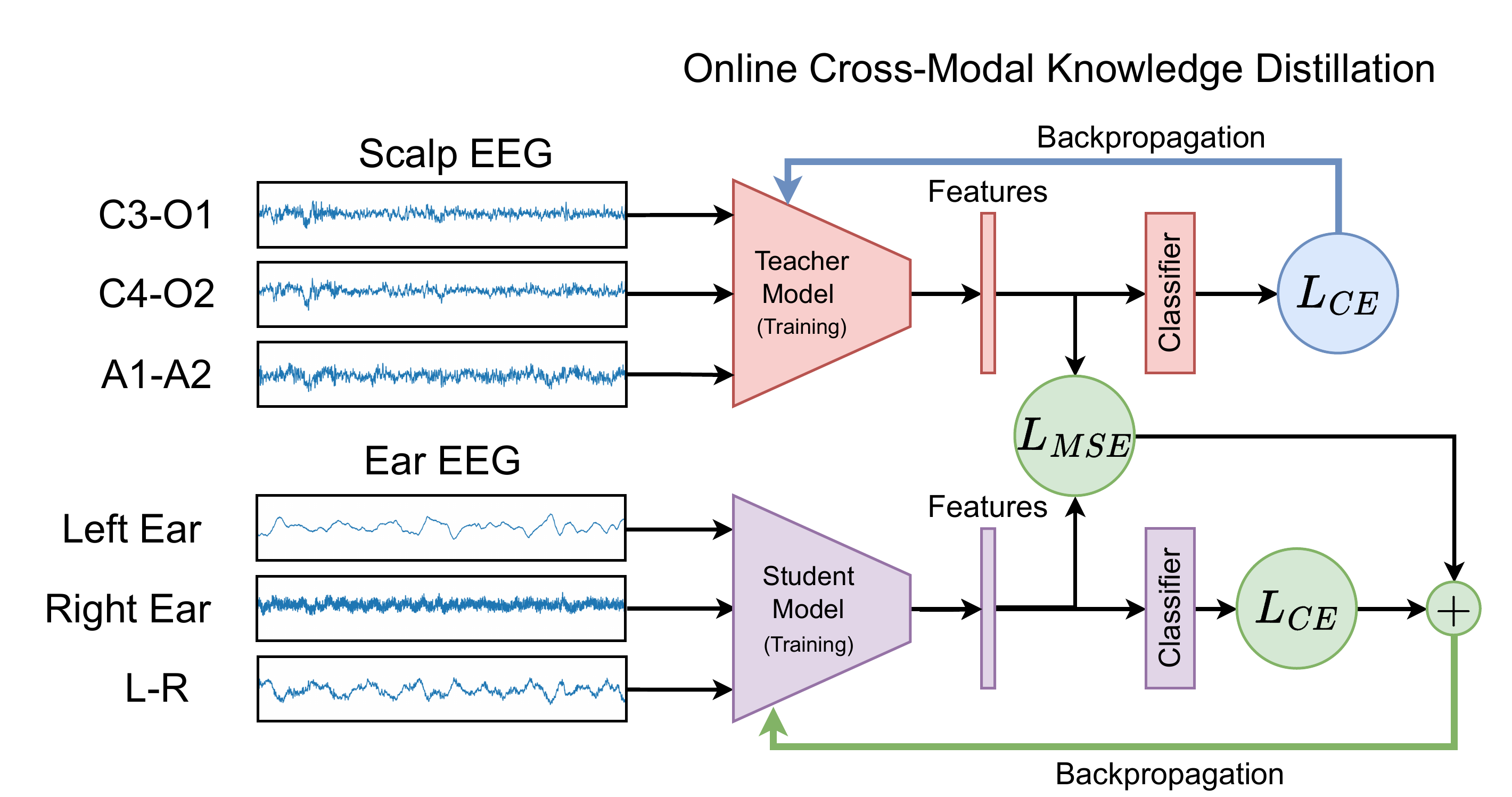}
    \vspace{-1em}
    \caption{Online cross-modal knowledge distillation framework.}
    \label{fig:crossmodalKD}
    \vspace{-1em}
\end{figure}

\vspace{-0.9em}
\subsubsection{Offline Knowledge Distillation}
\label{subsubsec:offline}
In offline feature-based knowledge distillation as shown in Fig.~\ref{fig:offline}, the teacher model $f^{T}_\theta$ is initially trained in the $\mathcal{X}^s$ domain in a supervised setting. In the offline knowledge distillation settings, the weights of $f^{T}_\theta$ are fixed, and the student model $f^{S}_\theta$ is randomly initialized and trained in the $\mathcal{X}^t$ domain by minimizing $\mathcal{L}^S_{kd}$, which is defined as:

\vspace{-0.25cm}
{\small
\begin{equation}\label{eq:1}
\begin{aligned}
    \mathcal{L}^S_{kd}(f^{S}_\theta(x^t_i),f^{T}_\theta(x^s_i), y_i) = \mathcal{L}^S_{ce}(f^{S}_\theta(x^t_i), y_i) \\
    + \mathcal{L}^S_{mse}(f^{S}_\theta(x^t_i), f^{T}_\theta(x^s_i)).
\end{aligned}
\vspace{-0.2cm}
\end{equation}}
Here $\mathcal{L}^S_{ce}$ represents categorical cross-entropy loss. $\mathcal{L}^S_{mse}$ is employed between the intermediate features of $f^{S}_\theta$ and $f^{T}_\theta$ to enable $f^{S}_\theta$ to learn a similar feature representation as $f^{T}_\theta$.

\vspace{-0.4cm}
{\small
\begin{equation}
    \mathcal{L}^S_{mse}(f^{S}_\theta(x^t_i), f^{T}_\theta(x^s_i)) = \frac{1}{N}\sum_{j=1}^{N}  \vert\vert f^{S}_{\theta}(x_{i}^{t}) - f^{T}_{\theta}(x_{i}^{s})\vert\vert_2^{2}.
\vspace{-0.3cm}
\end{equation}}
For clarity, $f_{\theta}(x_{i})$ is also used to denote intermediate feature maps.

\vspace{-0.9em}
\subsubsection{Online Knowledge Distillation}
In contrast to offline knowledge distillation, the weights of both $f^{T}_\theta$ and $f^{S}_\theta$ are randomly initialized and updated simultaneously in the online knowledge distillation setting as shown in Fig.~\ref{fig:crossmodalKD}. $f^{S}_\theta$ is trained to minimize the loss mentioned in 
\eqref{eq:1}, and $f^{T}_\theta$ is trained to minimize $\mathcal{L}^T_{ce}(f^{T}_\theta(x^s_i), y_i)$, which is a categorical cross-entropy loss. The goal is to improve ear-EEG based sleep staging by enabling $f^{T}_\theta$ and $f^{S}_\theta$ to jointly learn a common feature space between $\mathcal{X}^s$ and $\mathcal{X}^t$.



\vspace{-0.9em}
\subsection{Transfer Learning}
\label{subsec:TF}
Transfer learning is a deep learning procedure used for storing knowledge gained on solving a particular task and reuse them as an initialization point for a similar new task. This method has been widely used in sleep stage classification domain \cite{phan2020towards,phan2022sleeptransformer}, where a model is pre-trained on larger datasets and then fine-tuned towards smaller datasets. In our work, we employed transfer learning to improve ear-EEG based sleep staging and compared its performance with the knowledge distillation approaches. Here, we hypothesized that the model would learn the characteristics of scalp-EEG signals during pre-training and then adapts the weights toward ear-EEG during fine-tuning.



\vspace{-1.5em}
\section{Experiments}
\label{sec:experiments}

\subsection{Dataset}   
We used the dataset from \cite{mikkelsen2017automatic}, consisting of one whole night of sleep recording from 9 healthy subjects (age: 26-44; 3 females and 6 males). The dataset contains simultaneously recorded PSG and ear-EEG. The PSG consists of 8 scalp-EEG channels (O1, O2, C3, C4, A1, A2, F3, and F4 according to the international 10-20 system), 2 EOG channels, and a chin EMG channel. Note that the EOG and EMG channels were not considered in our study. The ear-EEG consists of 12 channels, with 6 channels from each ear: ELA, ELB, ELE, ELI, ELG, ELK, ERA, ERB, ERE, ERI, ERG, and ERK according to \cite{mikkelsen2017automatic}. All the EEGs were sampled at 200~Hz. The dataset was manually scored based on the international AASM \cite{Iber2007} guideline.

\vspace{-0.9em}
\subsection{Data Preprocessing}
All the channels (scalp and ear-EEG) were bandpass filtered between $0.2-42$ Hz. Noisy ear-EEG channels were removed based on the mean power. Initially, all ear derivations were calculated, and the mean power ($P_{ij}$) for derivation consisting of channel $i$ and $j$ within the frequency range of $10-35$ Hz was calculated \cite{mikkelsen2017automatic}. 
A channel $i$ was rejected if $m_i$ is an outlier, where $m_i$ = median($P_{ij}, \forall j$). Altogether 15 channels were rejected from the dataset. Recordings of subject $5$ were removed from the dataset as both the ear canal channels (ERA and ERB) from the right ear were rejected, thus average electric potential difference cannot be calculated. Table \ref{tab:rejected} shows the number of rejected and usable ear-EEG channels that remained after the channel rejection. As inputs to the network, derivations of scalp-EEG and ear-EEG were extracted as follows:\\
\textbf{Scalp-EEG : } Three scalp-EEG derivations were extracted from 8 scalp-EEG channels: C3-O1, C4-O2, and A1-A2. These three scalp-EEG derivations were chosen, because of their spatial correspondence with ear-EEG derivations.
\\
\textbf{Ear-EEG :} Three ear-EEG derivations were extracted from the 12 ear-EEG channels~\cite{mikkelsen2017automatic} as shown below, where $(\overline{\cdot})$ denotes the mean.

{\small
\vspace{-0.7cm}
\begin{align}
L_1 &= \overline{ELA, ELB, ELE, ELI, ELG, ELK} \\
R_1 &= \overline{ERA, ERB, ERE, ERI, ERG, ERK} \\
L-R &= L_1-R_1 \\
LE &= \overline{ELA, ELB} - \overline{ELE, ELI, ELG, ELK} \\
RE &= \overline{ERA, ERB} -  \overline{ERE, ERI, ERG, ERK}.
\end{align}}
\vspace{-0.7cm}

Here, $L-R$ derivation gives the average electric potential difference between the left and right ears. $LE$ and $RE$ derivations are the average potential difference between the concha and ear canal in the left and right ears, respectively.


\begin{table}[t!]
\vspace{-0.25cm}
\caption{Description of the rejected ear-EEG channels}
\label{tab:rejected}
\begin{center}

\scalebox{0.85}{

\begin{tabular}{|lccccccccc|}
\hline
\small{\bf Subject No} & \small{\bf \#1} & \small{\bf \#2} & \small{\bf \#3} & \small{\bf \#4} & \small{\bf \#5} & \small{\bf \#6} & \small{\bf \#7} & \small{\bf \#8} & \small{\bf \#9}\\
\hline
\small{Rejected Channels} & 2 & 2 & 1 & 2 & 5 & 0 & 0 & 1 & 2 \\
\small{Usable electrodes} & 10 & 10 & 11 & 10 & 7 & 12 & 12 & 11 & 10 \\
\hline
\end{tabular}}
\end{center}
\vspace{-0.2cm}
\footnotesize{*Recordings of subject 5 were not included in the study}\vspace{-0.65cm}

\end{table}

\vspace{-0.8em}

\subsection{Training and Experimental Setup}
The model under all training settings were trained using the Adam optimizer with learning rate ($lr$), $beta$1 and $beta$2 set to, $10^{-3}$, $0.9$ and $0.999$, respectively for $100$ epochs. The batch size was experimentally chosen as $32$.  The dataset was partitioned in a leave-one-subject-out (LOSO) fashion, as this is the most probable real-application scenario. Hence, at a given iteration, the model was trained on 7 subjects and tested on one, which the model has not seen during training. The model was implemented in PyTorch environment.




\vspace{-0.265cm}
\section{Results and Discussion}
We compared both our knowledge distillation and transfer learning strategies with the existing supervised learning architectures used in the domain of sleep stage classification. In order to show the robustness of the method, convolutional neural network (CNN) based U-Sleep\cite{usleep} and transformer based epoch cross-modal transformer \cite{pradeepkumar2022towards} were selected. Accuracy (ACC) and Cohen's kappa coefficient ($\kappa$) were considered as performance metrics in LOSO fashion.

\begin{table}[t]

\centering
\caption{\bf \small Overall performance of supervised learning, transfer learning (TL), and knowledge distillation (KD).}
\scalebox{0.85}{
\begin{tabular}{c|c|c|c|c}
\hline  

{\textbf{ Model }}  & \multirow{2}{*}{\textbf{Modality}} & \multirow{2}{*}{\textbf{Method}}  & \multirow{2}{*}{\textbf{ACC} }& \multirow{2}{*}{\textbf{$\kappa$}} \\

 \textbf{Architecture} &&&&\\
\hline
 
 \multirow{5}{*}{U-Sleep\cite{usleep}}   & Scalp-EEG &  Supervised & $73.19$  & $0.631$   \\
 \cline{2-5}
   & \multirow{4}{*}{Ear-EEG} &  Supervised & $64.58$ & $0.524$  \\
  & & TL    & $65.8$ & $0.533$\\
  & & KD (Offline)  &  $66.25$ & $0.544$  \\   
  & &  KD (Online)   & {\bf 68.04} & {\bf 0.562}  \\
 \hline

  & Scalp-EEG &  Supervised & $78.11$ & $0.670$  \\
 \cline{2-5}
   Epoch & \multirow{4}{*}{Ear-EEG} &  Supervised & $67.28$ & $0.532$  \\
  Cross-Modal & & TL & $63.84$ & $0.481$ \\
 Transformer & & KD (Offline) & {\bf 69.61} & {\bf 0.551} \\   
  \cite{pradeepkumar2022towards}& &  KD (Online) & $69.19$ & $0.551$  \\
 \hline

\end{tabular}}
\vspace{-2em}
\label{table:results}
\end{table}


 \begin{table}[!t]
\centering
\caption{{\bf \small Class-wise  MF1 performance of supervised learning, transfer learning (TL), and knowledge distillation (KD)}}
\scalebox{0.85}{
\begin{tabular}{c|c|c|c|c|c}
\hline  

\multirow{2}{*}{\textbf{\small{Method}}}  & \multicolumn{5}{c}{\textbf{\small{Per-class Performance}}} \\ \cline{2-6}

& \small{W}& \small{N1} & \small{N2} & \small{N3} & \small{REM} \\
\hline

\multicolumn{6}{c}{U-Sleep\cite{usleep}}\\
\hline

\small{Scalp-EEG$^{*}$} & $82.1 $ & $6.0 $ & $88.7 $ & $79.3 $ & $74.1 $ \\
\small{Ear-EEG$^{*}$} & $62.2 $ & $11.5 $ & $72.6 $ & $74.0$ & $60.0 $ \\


\small{TL} & $65.1\color{green}\uparrow$ & $6.3\color{red}\downarrow$ & $75.3\color{green}\uparrow$ & $74.6\color{green}\uparrow$ & $57.3\color{red}\downarrow$ \\


\small{KD (Offline)} & $62.2\color{green}\uparrow$ & $10.3\color{red}\downarrow$ & $74.7\color{green}\uparrow$ & $76.8\color{green}\uparrow$ & $60.8\color{green}\uparrow$ \\

 
\small{KD (Online)} & $64.3\color{green}\uparrow$ & $11.0\color{red}\downarrow$ & $77.5\color{green}\uparrow$ & $76.5\color{green}\uparrow$ & $61.6\color{green}\downarrow$ \\


\hline

\multicolumn{6}{c}{Epoch Cross-Modal Transformer\cite{pradeepkumar2022towards}}\\
\hline
 
\small{Scalp-EEG$^{*}$} & $83.9$ & $6.3$ & $88.7$ & $70.1$ & $79.4$ \\

\small{Ear-EEG$^{*}$} & $64.9$ & $6.5$ & $80.2$ & $70.3$ & $60.4$ \\

\small{TL} & $60.5\color{red}\downarrow$ & $2.2\color{red}\downarrow$ & $71.9\color{red}\downarrow$ & $69.2\color{red}\downarrow$ & $49.6\color{red}\downarrow$ \\


\small{KD (Offline)} & $60.6\color{red}\downarrow$ & $2.0\color{red}\downarrow$ & $85.9\color{green}\uparrow$ & $66.3\color{red}\downarrow$ & $66.5\color{green}\uparrow$ \\
 
\small{KD (Online)} & $68.4\color{green}\uparrow$ & $3.0\color{red}\downarrow$ & $81.8\color{green}\uparrow$ & $75.8\color{green}\uparrow$ & $58.3\color{red}\downarrow$ \\

\hline
\end{tabular}}

\raggedright
\footnotesize{$^{*}$ denotes supervised learning. $\color{green}{\uparrow}$ and $\color{red}{\downarrow}$ indicates performance increase and decrease compared to Ear-EEG$^{*}$.} 
\vspace{-0.65cm}
\label{table:class_per}
\end{table}
\noindent \textbf{Sleep Staging Performance:} Overall performance and comparison of the explored training strategies are given in Table \ref{table:results}. The results clearly state the effectiveness of both knowledge distillation training strategies in improving ear-EEG based sleep staging. For U-Sleep\cite{usleep}, the accuracy of the sleep staging was increased by $3.46\%$ and in Cohen's kappa coefficient by a margin of $0.038$, when compared to the supervised training. A similar performance increase was observed with epoch cross-modal transformers\cite{pradeepkumar2022towards}, which showed that the proposed knowledge distillation strategies were agnostic to model architectures. The Cohen's kappa coefficients obtained for supervised learning were low, when compared to more recent ear-EEG based sleep studies \cite{mikkelsen2019accurate}, but higher than the results presented in \cite{mikkelsen2017automatic}, where the dataset we used was originally presented. This lower Cohen's kappa coefficients could be attributed to a combination of several factors, including a lower sample size when compared to more recent studies \cite{mikkelsen2019accurate} and advanced equipment used to record ear-EEG in more recent studies \cite{mikkelsen2019accurate, Simondrycontact}.


Both online and offline knowledge distillation methods yielded similar scores. Considering the similar scores, offline knowledge distillation was preferred, because online knowledge distillation requires large computational resources as both teacher and student models were trained simultaneously. According to Table \ref{table:results} the proposed knowledge distillation strategies outperformed transfer learning in both model architectures. This indicates that the pre-trained model on the scalp-EEG domain was a poor initialization for the ear-EEG domain. 


Class-wise performances using macro averaged F1-score (MF1) of the training strategies are given in Table \ref{table:class_per}. In most sleep stages, the knowledge distillation methods improved the prediction, but it under performed predicting the N1 sleep stage, when compared to supervised learning. Generally, the Cohen's kappa for classification of the N1 sleep stage was low for all classification methods given in Table \ref{table:class_per}. A low Cohen's kappa for classification of N1 is a common trend for automatic sleep staging algorithms \cite{usleep,pradeepkumar2022towards} and therefore not a concern.




\begin{figure}[t!]
    \centering
    \includegraphics[width = 0.98\linewidth]{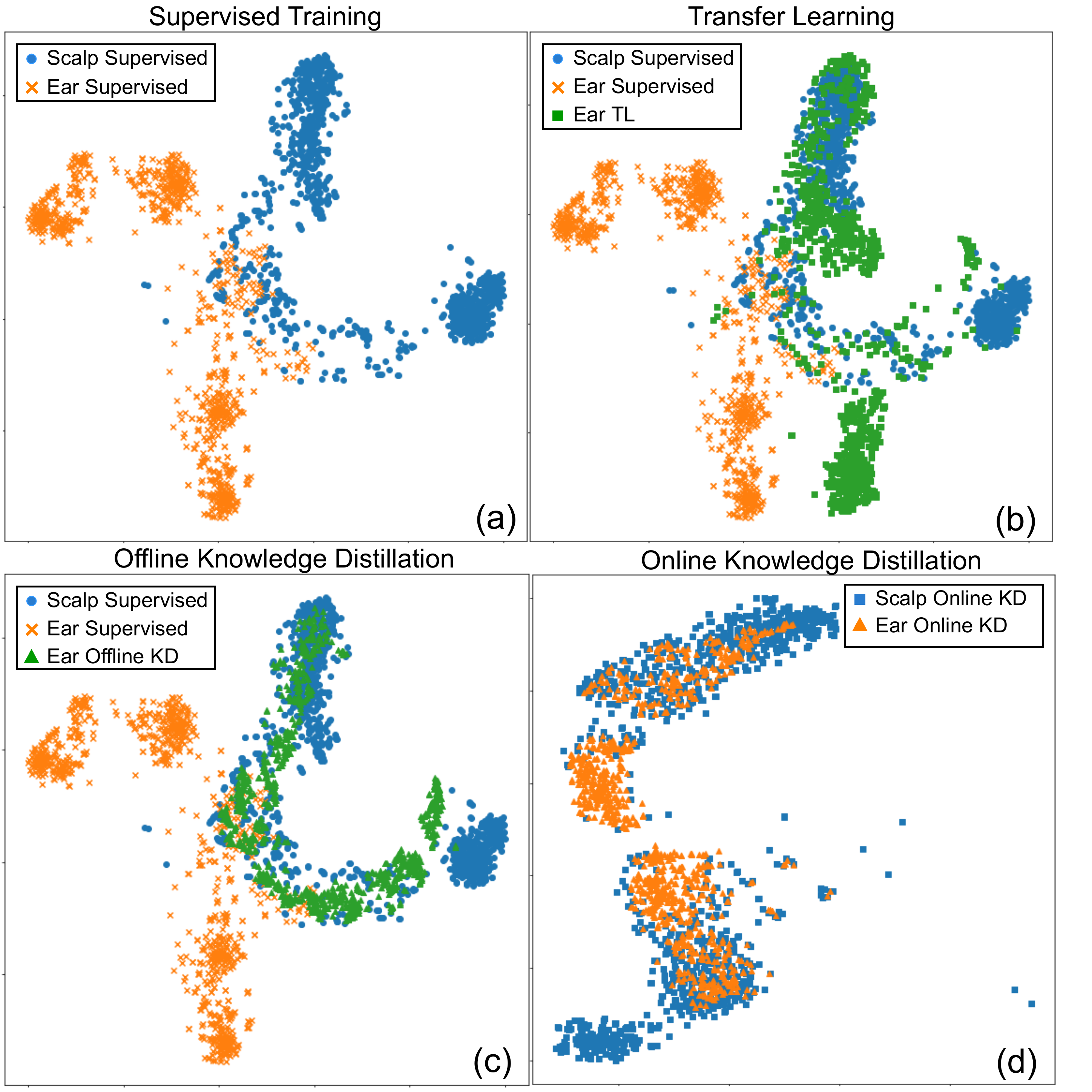}
    \vspace{-0.4cm}
    \caption{Distribution of features learned by U-Sleep. The distributions of features learned in supervised learning, transfer learning, offline knowledge distillation, and online knowledge distillation are given in (a), (b), (c), and (d) respectively. }
    \label{fig:tsne}
    \vspace{-0.3cm}
\end{figure}

\noindent \textbf{Analysis on Learned Representations:} In order to analyze and validate the learned representations by the student model under different training strategies, we extracted the features from intermediate layers and conducted T-distributed stochastic neighbor embedding (T-SNE). The visualizations of their distributions are shown in Fig. \ref{fig:tsne}. The plots clearly show two different clusters of scalp-EEG and ear-EEG features from the supervised trained models (Fig. \ref{fig:tsne}(a)). Fig. \ref{fig:tsne}(c) shows that the offline knowledge distillation strategy learns better features, which were similar to scalp-EEG, when compared to transfer learning (Fig. \ref{fig:tsne}(b)). Similarly, online knowledge distillation (Fig. \ref{fig:tsne}(d)) confirms our hypothesis on jointly learning a common feature space for better sleep staging.

\vspace{-1em}
\section{Conclusion}
\label{sec:conclusion}

\vspace{-1em}

In this paper, we present a domain adaption based training strategy to improve ear-EEG based sleep staging. We validated that the knowledge from scalp-EEG can be leveraged to enhance ear-EEG based sleep staging. We believe that this work opens the door for future research on unpaired domain adaptation techniques to utilize largely available scalp-EEG datasets and to analyze largely unexplored deep learning strategies to improve and achieve clinical standards in ear-EEG based sleep studies.  

\newpage



\section{Acknowledgement}

We would like to extend our gratitude to Center for Ear-EEG, Aarhus University, Denmark for proving us with the ear-EEG dataset. We also would like to thank Dhinesh Suntharalingham and Vinith Kugathasan for their support.

\small
\bibliographystyle{IEEEtran}
\bibliography{main}

\end{document}